\documentclass{article}

\usepackage{PRIMEarxiv}

\usepackage[utf8]{inputenc} 
\usepackage[T1]{fontenc}    
\usepackage{hyperref}       
\usepackage{url}            
\usepackage{xcolor}
\usepackage{booktabs}       
\usepackage{amsfonts}       
\usepackage{nicefrac}       
\usepackage{microtype}      
\usepackage{amsmath}
\usepackage{fancyhdr}       
\usepackage{graphicx}       
\graphicspath{{media/}}     
\definecolor{mygreen}{rgb}{0,0.5,0.2}
\title{Icarus’s Orbit as the Ultimate Test for the Vulcan Hypothesis and General Relativity}

\author{
	Pogossian S.P.\\
	Univ. Brest, CNRS, IRD, Ifremer, IUEM, Laboratoire d'Oc\'eanographie Physique et Spatiale (LOPS),\\ Technopôle Brest-Iroise, Rue Dumont d'Urville, 29280 Plouzan\'e \\
	Universit\'e de Bretagne Occidentale \\
	Brest, France\\
	\texttt{pogossia@univ-brest.fr} \\
}

\begin{document}
	\maketitle

	\begin{abstract}
		\ General Relativity quantitatively accounts for the anomalous perihelion precession observed in several planets’ orbits, with Mercury exhibiting the most significant deviation from Newtonian predictions. This anomalous advance, amounting to approximately 43 arcseconds per century for Mercury, was historically unexplained by classical mechanics, prompting hypotheses such as the existence of an intra-Mercurial planet, Vulcan. Although Vulcan was never observed, the concept persisted, and contemporary theoretical work has revived interest by proposing that this anomaly might instead be attributed to a compact massive object, such as a primordial black hole, within the inner solar system. Nonetheless, the successful prediction and measurement of Mercury's perihelion advance remains the cornerstone of the empirical validation of general relativity. In the present work, I carry out an in-depth study comparing methods for calculating perihelion advance of inner planets using two methods, one based on the rotation of the Laplace-Runge-Lentz vector and the other on the evolution of the perihelion longitude. I show that, although the two methods predict almost identical perihelion advances for inner planets according to classical gravitational theory and the general relativity model, they give divergent results for the asteroid Icarus. Comparison of numerical calculations of Icarus' perihelion advance with the prediction of the same advance by Einstein's formula leads to the conclusion that only the method based on Laplace-Runge-Lentz vector rotation provides a coherent explanation of the behavior of Icarus' perihelion advance.
		I then include the hypothetical planet Vulcan in the Newtonian gravitational model of the solar system, and analyze Vulcan's influence on the perihelion advance of the inner planets, using a set of Vulcan parameters obtained by optimization. Analysis of the high amplitude fluctuations in the perihelion advance evolution slope for the planet Venus allows us to exclude it from the list of inner planets. By selectively choosing Vulcan's mass as well as its position and other orbital parameters, the constraints of optimization enable us to obtain a very subtle agreement between Vulcan's semi-major axis and mass, necessary to simultaneously obtain the observed perihelion advances of Mercury, Earth and Mars. A hypothetical planet Vulcan, positioned between the orbits of Mercury and Venus with a semi-major axis a=0.545 AU and a mass approximately three times smaller than Mercury's, could exert gravitational influences on Earth and Mars comparable to those predicted by General Relativity. However, this Vulcan would have a significantly more pronounced impact on the perihelion advance of Icarus, inducing an effect about nine times stronger than the relativistic one. Therefore, precise observational data of Icarus's orbital dynamics offers a definitive means to either rule out this specific Vulcan hypothesis or, conversely, indicate the necessity of further refinement of General Relativity's application within the Solar System.
		
	\end{abstract}


	\section{Introduction}
	\
In astrodynamics, the accuracy of computational methods becomes particularly decisive when analyzing the dynamic trajectories of asteroids, especially those whose orbital parameters undergo significant variations. The ability to detect and analyze the specific characteristics of asteroid orbits provides valuable insights into their future evolution, which is especially crucial for Near-Earth Objects (NEOs), as accurate orbital determination allows realistic short-term predictions of potential impacts with Earth, the Moon, other planets, or artificial satellites. The uncertainties associated with orbital elements can have a substantial impact on long-term forecasts, making it imperative to use highly accurate computational models to assess potential collision risks. This is justified by the intrinsically chaotic nature of celestial mechanics, where small uncertainties in initial conditions can lead to large deviations in time due to the sensitivity of dynamical systems to such perturbations.  For asteroids with rapidly varying orbital elements such as those influenced by gravitational interactions with planets, non-gravitational forces like the Yarkovsky effect, or close encounters with massive bodies the propagation of uncertainties can quickly degrade the reliability of trajectory predictions. For near-Earth asteroids, which pass close to Earth's orbit, even minor errors in the calculated orbital elements can lead to significantly different trajectories in the near future. Given the catastrophic consequences of a potential impact, accurate and robust calculation techniques are essential to minimize prediction errors and provide reliable assessments of collision risks.  Furthermore, the ability to predict close approaches or impacts with other celestial bodies or artificial satellites in the vicinity of the Earth is essential for space situational awareness, planetary defense strategies and safeguarding operational spacecraft. 
	
Motivated by the need to improve the reliability of orbital predictions, I  conduct an in-depth review of some computational methods, with particular emphasis on the study of perihelion advance (PA) of planets and asteroid Icarus of our solar system, a key phenomenon for understanding orbital dynamics in complex gravitational systems. The calculation of this effect has historically served as a benchmark for evaluating the accuracy and robustness of dynamical models in celestial mechanics.  By examining the mathematical and physical frameworks employed to describe the anomalous PA, we gain valuable insights into the robustness of these methods, which can then be extended to refine our understanding of other complex orbital dynamics, including those of Near Earth asteroids (NEA) and NEOs.
Current astrodynamical modeling relies primarily on two gravitational frameworks: Newtonian mechanics, which remains effective for solar system-scale dynamics, and general relativity (GR), which accounts for relativistic effects such as spacetime curvature. 
Within the solar system, subtle deviations such as Mercury’s anomalous precession and precision spacecraft tracking underscore the necessity of relativistic corrections to Newtonian predictions. Beyond the solar system, galactic rotation curves exhibit velocities inconsistent with visible matter alone, motivating the widely studied dark matter hypothesis \cite{Carr2021}, \cite{Khlopov2010}. Meanwhile, certain astrophysical observations continue to challenge GR, prompting investigations into alternative gravitational models \cite{Milgrom1983}, \cite{Bekenstein2004}, \cite{Dvali2000}, \cite{Verlinde2011}.

It all began with Mercury’s anomalous motion; the 43 arcseconds-per-century ($^{\prime\prime}$/cy) precession of its perihelion marked the first definitive breach in Newton’s law of universal gravitation. Confronted with this challenge to classical physics, Urbain Le Verrier sought to preserve the Newtonian framework by postulating Vulcan, a hypothetical planet orbiting between Mercury and the Sun, whose gravitational pull would account for the discrepancy.
The unsuccessful search for the planet Vulcan over several decades by astronomers led Einstein to advance an explanation based on GR, which profoundly altered our conception of gravity and space-time. The solution to Einstein's equations is very complex (see \cite{Damour1985}), so an approximate approach was developed by Lorentz and Droste \cite{Lorentz_1_1917}, \cite{Lorentz1937}, \cite{Fienga2024} and later by Infeld, Hoffmann and Einstein himself, known as EIH equations \cite{EIH1938}, \cite{Infeld1960}. These equations, although approximate (and to my knowledge, no one has yet evaluated the error in the long-term integrations), have computationally affordable solutions only for the 2-body case (\cite{Damour1985}, \cite{Will2014_OK}).  According to Will\cite{Will2014_OK}, incorporating relativistic effects into an N-body system remains a highly complex task. The analytical solutions rapidly grow in complexity due to the proliferation of additional terms. At higher orders of the post-Newtonian expansion—such as the second (2PN) or third (3PN) order—these terms become so intricate that their physical interpretation is, in practice, exceedingly difficult, if not unintelligible \cite{Blanchet2001}.

The equations of GR used to describe the solar system are nonlinear and depend on the dynamical curvature of spacetime \cite{Lorentz_1_1917}, \cite{Lorentz1937}, \cite{EIH1938}, \cite{Turyshev2009}, \cite{Will2014_OK}, \cite{Poisson2014}, \cite{Newhall1983}, \cite{Standish1992_chpt8}.  Indeed, the two-body problem remains the only gravitational system that is well described within the framework of GR \cite{Damour1985}.

Due to the computational complexity of integrating post-Newtonian effects, in particular cross-interactions in N-body simulations, the Newtonian framework remains widely used when its accuracy is sufficient. In practical orbital mechanics, relativistic effects are often treated as small perturbations added to Newtonian gravity. While the first post-Newtonian approximation provides a common basis for such corrections, it still demands significant computational resources. In the solar system and similar weak-field regimes, where GR effects appear predominantly as small first-order corrections to Newtonian dynamics, a classical gravitational model incorporating dominant solar relativistic terms often suffices for practical astrophysical purposes \cite{Anderson1975},\cite{Quinn1991}, \cite{Benitez2008}, \cite{Hees2012}.

In Newtonian celestial mechanics, the solar system's barycenter is often used as a more practical frame of reference than a purely heliocentric one. While the barycenter follows an approximately inertial trajectory, experiencing only minimal acceleration relative to distant stellar frames,the Sun-centered frame undergoes measurable accelerations due to planetary gravitational interactions. These non-inertial effects can accumulate over time, making the barycentric frame essential for high-precision, long-term orbital integrations.\\ 
In the relativistic case, the masses of the moving bodies depend on the velocities of the bodies, and consequently the position of the barycenter, which performs a complex motion in space, is a function of the positions and velocities of the N bodies, and in principle constitutes an erratically accelerated reference frame, subject to further perturbations due to relativistic gravitational effects. The equations governing the barycentre are interdependent, requiring iterative solutions to determine the Sun's position and velocity. As mentioned by Fienga et al\cite{Fienga2024}, to handle the computational complexity of this problem, the relativistic barycenter equations are only solved at the initial stage of planetary integration. Subsequently, the equations of motion of the solar system bodies and the Sun are integrated within this fixed initial, relativistic barycentric frame.  Although this approximation simplifies the calculations, it introduces potential errors that may impact the reliability of orbital predictions. Such errors must be rigorously analyzed, especially in the context of long-term integration or dynamically evolving orbits, to ensure the accuracy of orbital forecasting.
Benitez and Gallardo in a paper fixed an objective to assess the significance of relativistic corrections in modeling the orbital dynamics of bodies within the Solar System, determining whether these corrections play a crucial role\cite{Benitez2008}. The planetary system model by Benitez and Gallardo includes the Sun and the eight major planets, from Mercury to Neptune, treating the Earth–Moon system as a single body positioned at its barycenter. They employed a simplified model of GR, originally proposed by Anderson et al\cite{Anderson1975} and further applied by Quinn et al\cite{Quinn1991} and Will\cite{Will2014_OK}, to describe the dynamics of the solar system. In this approximation, only the relativistic effects induced by the Sun were taken into account, and the motion of bodies took place in a relativistic space-time. Their results lead to the conclusion that while relativistic corrections are not essential for all dynamical studies, they are crucial for accurately modeling the motion of bodies in the inner Solar System.

In prior works, I evaluated the viability of the Vulcan hypothesis by analyzing orbital stability and physical plausibility, demonstrating that its mass and orbital parameters can be chosen to reproduce an anomalous PA of $43^{\prime\prime}$/cy for Mercury, while keeping the PA for Earth and Mars within a reasonable range of the values predicted by GR. I also explored the possibility that Vulcan could be a primordial black hole of approximately 1.6 Mercury masses, with a Schwarzschild radius of about 0.7 mm, rendering it invisible while still exerting sufficient gravitational influence to explain the anomaly, and maintaining orbital stability over the age of the solar system.
Although never directly observed, the Vulcan hypothesis, whether imagined as a classical planet or a primordial black hole, has practical value as a theoretical construct within Newtonian gravitation. Its significance lies not so much in the nature of the hypothesized celestial body but in its usefulness in improving the accuracy of orbital calculations within the framework of classical space and time. This concept allowed for the explanation of the PA of inner planets without significantly increasing computational complexity or introducing numerically expensive terms. Consequently, it enabled the use of straightforward, computationally efficient codes based on classical principles, effectively describing the behavior of the major planets in the solar system as an alternative to the simplified GR model \cite{Will2014_OK}, \cite{Quinn1991}, \cite{Anderson1975}, \cite{Newhall1983}.

In the present work, I compare the PA of the inner planets predicted by classical Newtonian gravity, with and without Vulcan, with those given by the simplified relativistic model for solar system proposed by Anderson et al\cite{Anderson1975} and used by a number of scientists \cite{Quinn1991},\cite{Benitez2008}, and \cite{Will2014_OK}. To reproduce the GR values for these advances I use another set of optimized orbital elements of Vulcan that differs from those used in my previous works. The core innovation of this new optimization approach involves the precise calibration of Vulcan’s physical and orbital parameters, such that its gravitational influence accurately reproduces the perihelion precessions of Mercury, Mars, and Earth in quantitative agreement with the predictions of GR for the Solar System. Throughout this paper, the term “Earth” refers to the Earth-Moon barycenter.
With such an optimized set of Vulcan's orbital elements available, a major challenge can be addressed by directly comparing GR and the Vulcan hypothesis through the analysis of the anomalous PA of another celestial body whose orbit approaches the Sun closely, and is therefore sensitive to the Sun's relativistic effects. The case of Icarus is well known, since it is considered to have largest anomalous PA after Mercury in our Solar system \cite{Gilvarry1953}, \cite{Francise1965}, \cite{Shapiro1968},\cite{Lieske1969},\cite{Shapiro1971}, \cite{Mahapatra1999}, \cite{Benitez2008}, \cite{Wilhelm2014}, \cite{Deines2017} and \cite{Greenberg2017}.  

To compare these two approaches for the case of the asteroid Icarus, one has to know the classical value of the PA since GR terms can be evaluated theoretically by the Einstein’s formula. However, in most of old and recent literature describing the orbit of Icarus I did not found the total PA of Icarus with or without relativistic corrections \cite{Gilvarry1953}, \cite{Francise1965}, \cite{Shapiro1968},\cite{Lieske1969},\cite{Shapiro1971}, \cite{Mahapatra1999}, \cite{Benitez2008}, \cite{Wilhelm2014}, \cite{Deines2017} and \cite{Greenberg2017} since it has not been reported before maybe due to the very high eccentricity of it’s orbit.
In the present work I calculate the PA predicted by GR theory for this asteroid, and my results concord perfectly with the predictions calculated by GR term of PA reported in \cite{Null1968}, \cite{Lieske1969}, \cite{Shapiro1971}, \cite{Sitarski1992}, \cite{Mahapatra1999}, \cite{Benitez2008}, \cite{Wilhelm2014}, \cite{Deines2017} and \cite{Greenberg2017}. Most strikingly, the Vulcan theory gives a value of  $91.39^{\prime\prime}$/cy, which is about nine times greater than that predicted by GR. So the experimental measurement of  the value of Icarus’s PA can definitively invalidate the Vulcan hypothesis in comparison to GR theory. Thus, an experimental determination of the total PA, based solely on observational data, could serve as a definitive test of Anderson’s simplified relativity model for the solar system \cite{Anderson1975}.

	\section{Calculations}
	\label{sec:2}
The total PA of a celestial body can be determined through precise astronomical observations. This observed precession can be formally separated into two components, the classical contribution predicted by Newtonian gravity, and an additional relativistic correction arising from Einstein's theory of GR. The relativistic component of this precession can be independently calculated using Einstein's analytical expression for PA, given the orbital parameters of the celestial body \cite{Stewart2005}. 	

\begin{equation}
	\Delta \omega = \frac{6\pi \mu_{\text{Sun}}}{c^2 a (1 - e^2)}
\end{equation}

where \( \Delta \omega \) is the relativistic advance of the perihelion per orbit, \( \mu_{\text{Sun}} = G M_{\text{Sun}} \) is the Sun’s standard gravitational parameter, \( c \) is the speed of light in vacuum, \( a \) is the semi-major axis, and \( e \) is the orbital eccentricity.

The two components of PA, the classical (Newtonian) and the relativistic component, can be evaluated through two distinct and independent approaches. 
The first classical (Newtonian) contribution to the PA of a given celestial body is calculated using the classical equations of motion derived from Newton’s law of universal gravitation. Within this framework, alternative hypotheses such as the existence of the hypothetical planet Vulcan can also be tested.

The second relativistic component (predicted by GR) is evaluated by comparing the observed total PA with that obtained by the classical Newtonian contribution. Total PA can be calculated within the simplified GR framework described by Anderson et al \cite{Anderson1975}.
So an interesting test of numerical evaluation methods of PA lies in the comparison: the difference between the observed total PA and the calculated Newtonian PA should, within observational uncertainties, precisely equate to the value directly predicted by Einstein's relativistic formula Eq.(1). This methodology also provides a robust verification of the Vulcan hypothesis.
To ensure a relevant comparison between the predictions of classical Newtonian gravitational theory (with or without Vulcan) and GR about PA of a planet, it is essential to rigorously define the latter and the methods used to calculate it, paying particular attention to the initial conditions and the definition  of orbital  parameters in each theoretical framework. 

Let us return to the notion of perihelion, which is strictly defined only in the case of Keplerian two-body elliptical motion. In this idealized case, the orbit is assumed to be planar and closed, the force purely Newtonian, the orientation of the orbit remains fixed in space, and the period of revolution is constant in time. The perihelion is defined as the point on the elliptical orbit closest to one of its two foci. In planetary motion, the Sun is located at one of these foci, so the perihelion corresponds to the planet’s point of closest approach to the Sun during each revolution. The direction of the perihelion is determined by a conserved Laplace–Runge–Lenz (LRL) vector, which lies along the major axis of the ellipse and points from the central mass toward the perihelion. In Keplerian elliptical orbits, the perihelion and aphelion remain fixed in space, and the apsidal line connecting them passes through the center of attraction.
The definition of perihelion is less straightforward for the N-body problem, since orbits are no longer closed 2D curves but strictly speaking non-periodic 3D curves. The sidereal orbital period of a planet is usually defined as the time between two transits of the planet at the same point relative to the distant stars. In the case of N-body problem this definition must be slightly modified since the planets have evolving orbits under the influence of other planets attraction. Their trajectory is thus open and therefore the planets pass through different points at each revolution. In this case, one can only refer to a mean period of revolution, as defined in one of my previous works \cite{Pogossian2022}. The concept of perihelion, also, can be extended to open trajectories close to the ellipse and known at least on one mean period. Thus, the extended perihelion can be defined as the closest point to the Sun among all the other points of a quasi-elliptical trajectory of the same mean period.
Although not all of the trajectory's points are coplanar, a mean elliptical trajectory can also be associated to each planet's quasi-elliptical orbit around the Sun within one mean period. I have shown in a previous work that a mean plane and a mean ellipse can be fitted to the orbit of each planet for every revolution around the Sun. The major axis of the mean ellipse is taken as the apsidal line for each revolution. The geometrical perihelion is the point on the ellipse closest to the focus of the mean ellipse near which the Sun is positioned.
The advance of the extended perihelion and the geometrical perihelion can be characterized by the angle of rotation of the radius vectors pointing, respectively, to these perihelia with respect to a predefined fixed reference direction. In the aforementioned work, the LRL vector was also used to define a third definition of PA. In this approach, the rotation of the LRL vector relative to a fixed reference direction is interpreted as the PA. As shown in the aforementioned (\cite{Pogossian2022})  all these three definitions converge to Newtonian classical value of  Mercury’s PA, but  the PA defined by the LRL vector rotation have the smallest standard deviation. This is why all along this paper I will retain PA determination via LRL vector rotation with respect to a fixed reference direction.
In most contemporary studies\cite{Will2014_OK}, \cite{Park2017}, the perihelion advance of a planet is generally calculated by determining the average rate of change of its longitude at perihelion $\frac{d\varpi}{dt}$ : 

\begin{equation}
	\frac{d\varpi}{dt} = \frac{d\omega}{dt} + \cos i \, \frac{d\Omega}{dt}
\end{equation}

Here, \( \omega \) is the argument of perihelion and \( \Omega \) is the longitude of the ascending node. The argument of perihelion \( \omega \) of a body in the Solar System defines the orientation of the perihelion within the orbital plane, measured from the ascending node, which is the intersection of the orbital plane with the ecliptic. The longitude of perihelion \( \varpi \) describes the orientation of the perihelion in three-dimensional space. Its rate of change corresponds to the observable perihelion precession, which results from both the in-plane rotation, represented by changes in the argument of perihelion \( \omega \), and the nodal precession, represented by changes in the longitude of the ascending node \( \Omega \).
The longitude of the ascending node \( \Omega \) is measured in the ecliptic plane relative to a reference direction, usually the vernal equinox. This measurement reflects the combined effects of the nodal line’s motion, caused by orbital perturbations, and the drift of the reference direction, such as that resulting from the precession of the equinoxes. While the vernal equinox is commonly used as a reference direction in celestial mechanics, its position is affected by Earth’s axial precession primarily driven by gravitational torques from the Sun and Moon and does not reflect the intrinsic orbital motion of the considered planet.\\
The widespread use of osculating elements is due to their usefulness in the two-body problem, where only the central body and the orbiting body interact. In this idealized case, the Keplerian orbital elements remain constant and provide a clear geometric description of the orbital motion. This framework serves as a basic model for understanding the dynamics of celestial bodies under the influence of the central body’s gravitational field.  
However, in our solar system, various disturbing forces such as gravitational interactions with other celestial bodies, solar radiation pressure, and relativistic effects due to the Sun’s gravity induce gradual changes in the orbit’s shape and orientation, leading to deviations from the ideal Keplerian motion. Under these perturbations, the orbital elements cease to be constant and evolve gradually over time, reflecting the dynamic and complex nature of actual orbital mechanics.

The gradual changes in the orbits of celestial bodies due to many-body gravitational interactions can be described using two complementary approaches. One relies on the state vector, which specifies a body's position and velocity at a given instant. This vector evolves continuously over time according to Newton's laws and provides a direct and exact description of the orbital motion. Complementarily, the other approach employs osculating elements, which offer a geometric interpretation of the motion and are especially useful for analyzing how perturbing forces affect familiar orbital parameters. These elements define the instantaneous Keplerian orbit that would match the body’s position and velocity if all perturbations were momentarily removed. 

State vectors and osculating orbital elements are mathematically equivalent representations of an orbit, yet each is optimally suited for different applications in astrodynamics. The state vector formulation simplifies   the implementation of integration schemes, enabling accurate, robust, and efficient computation of orbital trajectories through direct integration of Newton’s laws, particularly in the presence of various perturbative forces.
Osculating orbital elements provide a geometrically intuitive representation of orbital motion. Their evolution is governed by the Lagrange planetary equations, which relate their time variation to the perturbing function \( R \). Since perturbative forces are considerably weaker than the dominant central gravitational force, the osculating orbital elements exhibit only small-amplitude variations over short timescales. This characteristic reveals subtle secular and periodic trends that are essential and analytically meaningful for understanding long-term orbital dynamics.
The perturbation function \( R \), often derived from physical force models expressed in Cartesian coordinates, must be reformulated in terms of osculating orbital elements to be compatible with the Lagrange planetary equations. In analytical treatments, this transformation is commonly achieved by expanding \( R \) as a power series in eccentricity. Such expansions converge only for eccentricities below  \( e \) <  0.6627, the so-called  Laplace limit. For practical purposes, truncated series are frequently used to calculate long-term variations in orbital elements caused by disturbing forces.

The time evolution of the argument of perihelion \( \omega \) , and the ascending node, \( \Omega \) , both are subject to complex behavior under the perturbing influence of other planets, including a gradual change in relative orientation of the ecliptic and orbital plane of the planet, which leads to a shifting of the nodal line and the advance of the perihelion. Furthermore, calculating the planetary PA from the time series of the longitude of perihelion involves calculating the projection of the ascending node \( \Omega \) onto the orbital plane of the planet. This projection also depends on the angle of inclination between the orbital plane and the ecliptic, and therefore the time evolution of the inclination will also affect the mean value of \( {d\varpi}/{dt} \). 

As mentioned above the perturbation function \( R \) is generally calculated using truncated series in the Lagrange planetary equations in terms of the orbital elements. While most studies focus on the long-term secular variations, high-frequency variations receive less attention. Secular terms account for slow variations occurring over timescales much longer than the orbital period, while high-frequency variations stem from small-amplitude changes and can lead to rapid oscillations caused by quasi periodic perturbations. These fast variations are often overlooked, possibly due to the complexity of the perturbation theory needed to analyze them. However, if abrupt changes occur in orbital elements such as during close encounters, resonances, or high eccentricity case, \( {d\varpi}/{dt} \) can exhibit sharp spikes. For example, in the case of Icarus, the very large eccentricity of Icarus’ orbit and the close planetary approaches make the analytical computation difficult.  In the case of Mercury, as pointed out by Narlikar and Rana \cite{Narlikar1985}, when Venus approaches Mercury at aphelion a rapid advance of Mercury's perihelion occurs. While this advance can reach up to $10^{\prime\prime}$ within a few weeks around the conjunction, it is important to note that the relativistic contribution averages only  $43^{\prime\prime}$/cy.  Notably, in simplified models of the solar system where the Earth's axial rotation is neglected and orbital planes of considered celestial bodies are assumed to be inclined only a few degrees with respect to the ecliptic plane, the determination of the LRL vector rotation and the classical computation of the mean rate of change of the longitude of perihelion over a 100-year interval both rely fundamentally on the dynamical evolution of the perihelion.  
My current calculations are performed within a state vector framework, utilizing either the classical Newtonian model \cite{Pogossian2022} or a simplified GR model \cite{Anderson1975}, \cite{Quinn1991}, \cite{Benitez2008}. After obtaining the state vector, the time series of orbital elements and the LRL vector are computed at each orbital position. Since this work builds upon previously developed simulations, we direct the reader to \cite{Pogossian2022}, \cite{Pogossian2023}, \cite{Pogossian2025} for a detailed description of the integration scheme and solver parameters utilized in this analysis.\\
First, I will compare two methods for determining the PA: the first based on the rotation of the LRL vector, and the second based on the mean rate of change of the longitude of perihelion over a 100-year period.  Subsequently, I will compare the PA values obtained by both methods using two different models: firstly, the Vulcan model \cite{LeVerrier1859}, \cite{Pogossian2023}, \cite{Pogossian2025}, and secondly, the GR model \cite{Anderson1975}, \cite{Quinn1991},\cite{Benitez2008}.\\
The first method for calculating the PA of the inner planets is based on the rotation of the LRL vector with respect to a fixed direction in space. This numerical technique extends the approach developed in \cite{Pogossian2022} and utilizes orbital data for the computation of PA. Among the three methods considered, the LRL vector rotation was selected due to its superior accuracy, yielding the lowest error compared to the two alternatives analyzed in \cite{Pogossian2022}. Next, the time evolution of the PA for the inner planets is modeled using a linear affine function over a 900-year period, independent of the specific method or model used to derive the PA values. This linear fit yields the mean rate of change (slope) of PA over the full interval, along with its standard error. These parameters are then used to estimate the expected variation in PA over a 100-year timescale, including the associated uncertainty.
The second method, which is the most widely used, determines PA by calculating the average change in the longitude of perihelion over a 100-year interval \cite{Will2014_OK}, \cite{Park2017}. To maintain consistency when comparing with our PA derived from the rotation of the LRL vector, the time series of the perihelion longitude, defined as $\varpi = \omega + \Omega \cos i$ \cite{Will2014_OK}, \cite{Park2017}, is fitted to a linear affine function to model its quasi-linear evolution over a 900-year interval, and the resulting slope is then used to estimate the corresponding change in perihelion longitude over a 100-year period.

The first solar system model considered in this work is based on the Newtonian classical gravitation equations with 8 planets (and 9 planets when Vulcan is included) in translation around the sun. The asteroid Icarus will be the 9-th (or 10th) body.  The Newtonian gravitational equations for a 10-body problem (11-body problem when Vulcan and Icarus are included) were integrated over a time interval of 333473 days (913 years). Jupiter and Saturn, being the most massive planets in the Solar System situated closest to the inner terrestrial planets, exert the dominant gravitational perturbations on the quasi-elliptical orbits of the latter. The orbits of these massive planets are locked by a mean motion resonance of 2:5, and about every 900 years, the direction of alignment of these giant planets returns almost exactly to its original direction  \cite{Wilson1985}, \cite{Pogossian2022}, \cite{Pogossian2023}. Therefore, I have chosen a time interval close to the period of \emph{Great Inequality} to consider all the possible relative positions of these giant planets with respect to the orbit of each of the inner planets.  As described in detail in my previous works \cite{Pogossian2022}, \cite{Pogossian2023}, \cite{Pogossian2025}, the 10-body solar system (11- body problem when the Sun, Vulcan and asteroid Icarus are included) has been integrated based on purely Newtonian initial conditions taken from DE200/LE200 ephemerides, subtracting corrections based on GR \cite{LeGuyader1993}, \cite{Arminjon2004}.

The second solar system model is a simplified relativistic model based on Quinn’s integration procedure for relativistic equations, initially proposed by Anderson et al\cite{Anderson1975} and detailed in the work of Quinn et al \cite{Quinn1991}. Following Benitez and Gallardo \cite{Benitez2008}, I include relativistic corrections to model the orbital dynamics of Solar System bodies. The system consists of the Sun and the eight major planets, with the Earth-Moon system treated as a single body at their barycenter. Another body like asteroid Icarus is added to the system, when 10 body problem is considered. In this above approximation only Sun’s induced GR effects are taken into account. For the sake of later comparison with the Vulcan model, I chose the same initial values for velocities, positions, masses, and other parameters of the solar system based on the DE200 ephemeris \cite{Standish1990},\cite{Standish1992_chpt8},\cite{Seidelmann1992}, but in this case I use initial conditions from the DE200 ephemeris including relativistic corrections and not the Le Guyader data bases \cite{LeGuyader1993}. 
The total observed PA of a celestial body can be formally represented as the sum of two gravitational components, a classical contribution predicted by Newtonian gravity and an additional relativistic precession predicted by GR.\\
The Newtonian contribution to the PA of the given celestial body can be calculated using the classical equations of motion and using Newton's law of universal gravitation.The GR component can be evaluated by two different independent ways. Before presenting the first method, let us recall that the observed total PA is the sum of classical component and GR components. The total PA can be calculated from the equations of motion of the celestial body by adding the relativistic corrections to the Newtonian equations, whose simplified form is described in \cite{Anderson1975}, \cite{Quinn1991},\cite{Benitez2008}. Therefore, the first method to calculate the relativistic component of PA is based on the evaluation of the difference between the total PA and the Newtonian component. The second method for determining the relativistic component of PA is based on analytical expression of Einstein for relativistic component of PA Eq.~(1) \cite{Einstein1915}, \cite{Stewart2005}, given the orbital parameters of the celestial body. This comparative methodology provides a systematic way to thoroughly assess the level of agreement between a specific numerical method for determining PA and various theoretical models.

Specifically, this critical test compares the difference between the observed total PA and its Newtonian component with the value predicted directly by Einstein's relativistic formula. Within observational uncertainties, these quantities should be equal. 
This methodology enables a comparative analysis of the two approaches for determining the PA introduced earlier: one relying on the rotation of the LRL vector, and the other on the variation of the perihelion longitude. The comparison emphasizes their respective differences and similarities. Both methods were assessed over a 100-year period using the same processing database.

\section{Results and Discussion}
\label{sec:3}
\
In the current study, within the framework of the Vulcan hypothesis, the orbital elements and mass of Vulcan were optimized to match the PA values predicted by GR for all inner planets except Venus. This optimization does not extend to Venus because its PA exhibits oscillations of very large amplitude, sometimes reaching significant magnitudes around an overall linear trend \cite{Pogossian2025}, and having very low coefficient of determination. This distinctive apsidal precession behavior of Venus, caused by the opposing gravitational influences of Earth and Mercury on one side and Jupiter and Mars on the other, has been thoroughly described in previous work \cite{Pogossian2025}. 

To describe the linear trend of planetary PA, a linear regression is performed, with the intercept and slope estimated by minimizing the sum of squared vertical deviations between the observed PA values and the fitted line. To evaluate the adequacy of the linear regression model, I treat deviations from the linear trend as if they were random, despite their origin in complex, deterministic Newtonian N-body  interactions (with or without GR correction terms). Specifically, I model the aggregate behavior of these deviations over the considered 900 year time interval as approximately normally distributed around the linear trend. Accordingly, for regression diagnostics, I assume that the residuals behave like a random, normally distributed error term. Under these assumptions the standard deviation of the slope in the linear trend was calculated, and the 95\% confidence interval (CI) was derived using the Student’s t-distribution, assuming a large sample size  $N \sim 8.0 \times 10^6$ \cite{Montgomery2018}.  

The PA values for the inner planets, summarized in Table-1 and calculated within the framework of classical Newtonian gravitation, are obtained using two distinct methods. The first method determines the angle of rotation of the LRL vector relative to a fixed direction in the initial orbital plane \cite{Pogossian2022}, \cite{Pogossian2023}, \cite{Pogossian2025}.  As demonstrated in the aforementioned studies, the rotation of the LRL vector exhibits a secular linear trend over time. Once the slope and its associated variance are determined over a 900-period interval, the PA is quantified as the average evolution of this rotation over a one-century timescale.
	
\begin{table*}[h]
	\centering
	\caption{This table presents the PA determined for each solar system body using two distinct methods: the rotation of the LRL vector and the time evolution of the longitude of perihelion ($\varpi$). Each method’s results are shown in separate columns, followed by the corresponding coefficient of determination ($\mathbf{R^2}$) reflecting the quality of the linear fit.}
	\label{tab:Table1}
	\begin{tabular}{|p{2cm}||p{2.5cm}|p{2.5cm}|p{2.5cm}|p{2.5cm}|} 
		\hline
		\multicolumn{5}{|c|}{\textbf{Table 1 : Newtonian component of PA}} \\
		\hline
		\textbf{Body} & \textit{PA (LRL)} in $^{\prime\prime}$/cy & $\mathbf{R^2}$ & \textit{PA} ($\varpi$) in $^{\prime\prime}$/cy & $\mathbf{R^2}$ \\
		\hline
		Mercury & $531.97 \pm 0.00$ & 0.99997 & $533.38 \pm 0.00$ & 0.99997 \\ 
		\hline
		Venus & $-35.42 \pm 0.18$ & 0.01783 & $-36.78 \pm 0.18$ & 0.01919 \\
		\hline
		Earth & $1156.86 \pm 0.08$  & 0.99016 & $1155.21 \pm 0.08$ & 0.99013 \\
		\hline
		Mars & $1591.68 \pm 0.04$  & 0.99853 & $1591.44 \pm 0.04$ & 0.99853 \\
		\hline
		Icarus & $248.48 \pm 0.01$  & 0.99243 & $968.43 \pm 0.02$ & 0.99911 \\ 
		\hline
	\end{tabular}
\end{table*}

In a separate column for each method of determining the PA, the coefficient of determination $\mathbf{R^2}$ associated with the linear fit is presented. Commonly denoted by $\mathbf{R^2}$, this statistical metric is used to assess the goodness-of-fit of a linear regression model and represents the ratio of the explained variance to the total variance.  
Although the estimated slope of $-35.42 \pm 0.18$ for the linear time evolution of Venus's PA (LRL) may appear statistically precise, this narrow CI reflects the large sample size  ($N \sim 8 \times 10^6$) rather than its scientific relevance. 
Additionally, the very low coefficient of determination ($\mathbf{R^2}< 0.018 $) indicates that the linear model explains less than 2\% of the variance in the data, suggesting that the trend accounts for only a minimal portion of the actual variability observed.
Therefore, in the case of Venus, the observed linear time evolution of its PA, despite its statistical precision, is not a scientifically convincing or robust representation of the phenomenon under study. As a result, it will be disregarded in scientific discussions.
This may explain why Venus's PA has often been omitted from various astronomical datasets, including those in Clemence’s seminal work\cite{Clemence1947}, which remains a cornerstone reference in the field.\\ 
For the other inner planets, the slope's 95\% confidence interval exhibits negligible relative uncertainty (\(< 7 \times 10^{-5}\)), indicating a statistically significant linear trend.  The very high coefficient of determination ($\mathbf{R^2 > 0.99}$) strongly supports the robust secular linear evolution of these bodies' PA, reflecting a long-term gravitationally governed trend that remains consistent over the $900$-year interval.
As established in a prior work\cite{Pogossian2025}, the PA of Mercury, Earth and Mars exhibits an approximately linear trend over 900-year intervals, modulated by low-amplitude seemingly aperiodic oscillations. 
The evolution of PA of the inner planets given  in Table-1 were carried out using classical Newtonian gravitation without incorporating the hypothetical planet Vulcan. The Newtonian value for Mercury's PA= $(531.97 \pm 0.00)^{\prime\prime}$/cy in Table-1, indicating the highest statistical significance among all solar system planets. This is confirmed by the highest value of the coefficient of determination, $\mathbf{R^2}$=0.99997. This indicates an almost perfect linear evolution of the PA over the 900-year time interval and further supports this approach. The regression analysis of PA of the Earth, Mars and Icarus reveals a linear time evolution with an estimated slope of $(1156.86 \pm 0.08)^{\prime\prime}$/cy, $(1591.68 \pm 0.04)^{\prime\prime}$/cy and $(248.48 \pm 0.01)^{\prime\prime}$/cy respectively with very high coefficient of determination $\mathbf{R^2}$ > 0.99  in the 900 year time interval. 
Another notable observation from Table-1 is that the PA for all planets calculated with LRL vector rotation approach and with the help of perihelion longitude ($\varpi$), within a 95\% CI, is nearly identical for all the inner planets, including Venus. This consistency, however, does not extend to the asteroid Icarus.\\
In Fig. 1, the evolution of PA based on the rotation of the LRL vector for Mercury, Venus and Earth are presented with the corresponding linear trends. From Table -1, it should be noted that their 95\% CI are so narrow that, in the scale shown, the empirical lines representing the upper and lower limits of the CI for these slopes of the LRL vector time series overlap. They are therefore not shown in the Fig. 1.

\begin{figure*}[!htb]
	\centering
	\resizebox{0.999\textwidth}{!}{\includegraphics[height=4.1 cm, width=7.2 cm,angle=0]{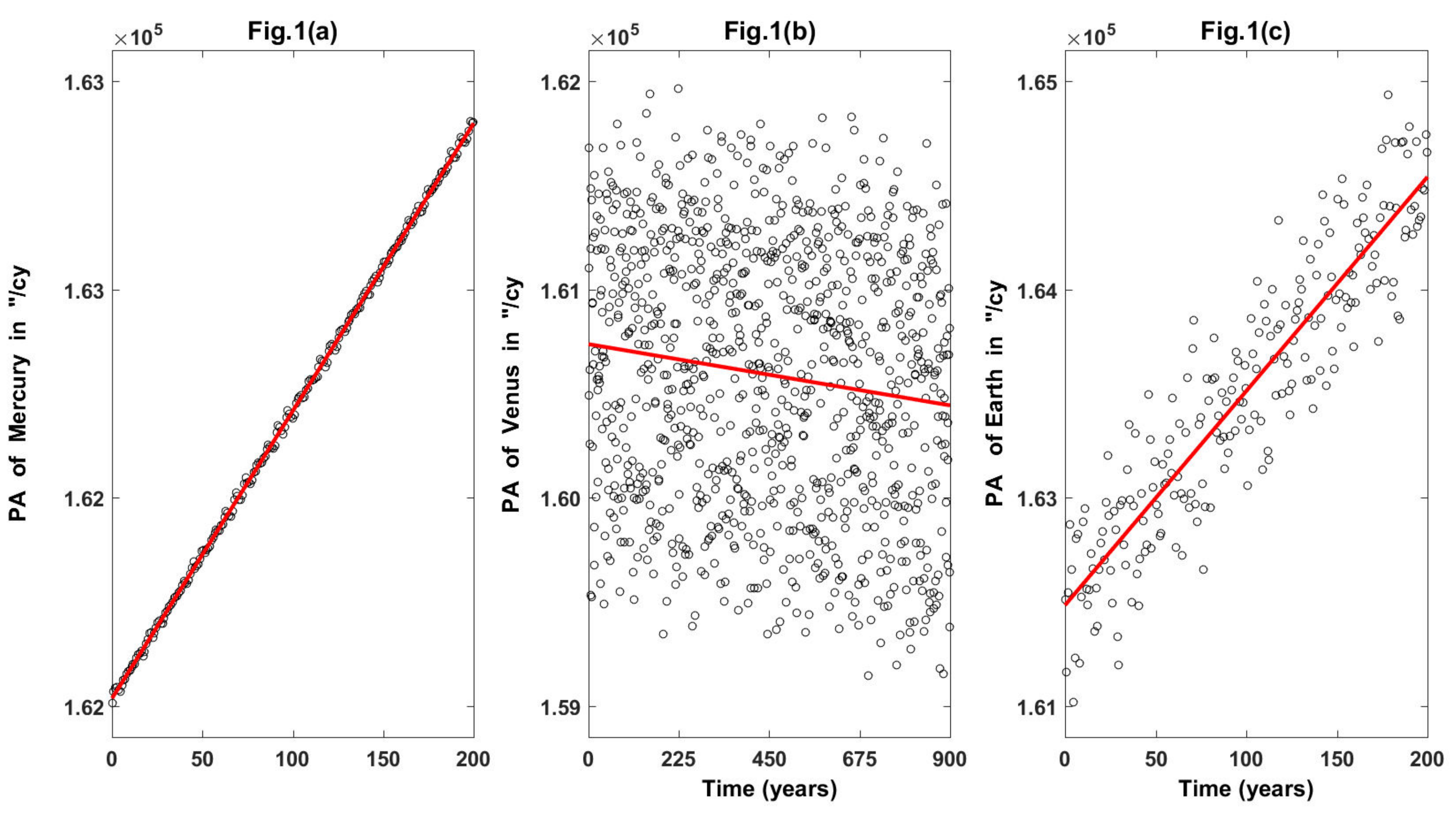}}
	\caption{\label{Fig. 1 :} 900-year evolution of PA of : (a) Mercury (b) Venus  and (c) Earth, with corresponding fitted linear trends.To enhance visual clarity and reduce rendering complexity, the original time series data, comprising approximately $8 \times 10^6$ points, were uniformly downsampled to 1000 points using equally spaced intervals, preserving the overall temporal structure and trends. Additionally, to facilitate more detailed visual analysis and focus on short-term variations, the time span shown in Fig. 1a and Fig. 1c was limited to a 200-year interval.
	}
\end{figure*}

I present in Fig. 2 comparative analysis of the time  evolution of the LRL vector's rotation and the longitude of perihelion ($\varpi = \omega + \cos(i) \times \Omega$) with respect to a fixed reference direction, using Mars as an example within the Newtonian gravitational framework. This analysis does not include the hypothetical planet Vulcan. Fig. 2a shows the rotation of the LRL vector and the longitude of perihelion over a 900-year interval, while Fig. 2b presents a 100-year zoomed-in view. Solid lines indicate the best-fit linear trends for the LRL vector. To clearly distinguish the closely spaced curves, the origin of the perihelion longitude curve has been shifted upward by an arbitrary value.\\ 

\begin{figure*}[!htb]
	\centering
	\resizebox{0.999\textwidth}{!}{\includegraphics[height=4.0 cm, width=7.2 cm,angle=0]{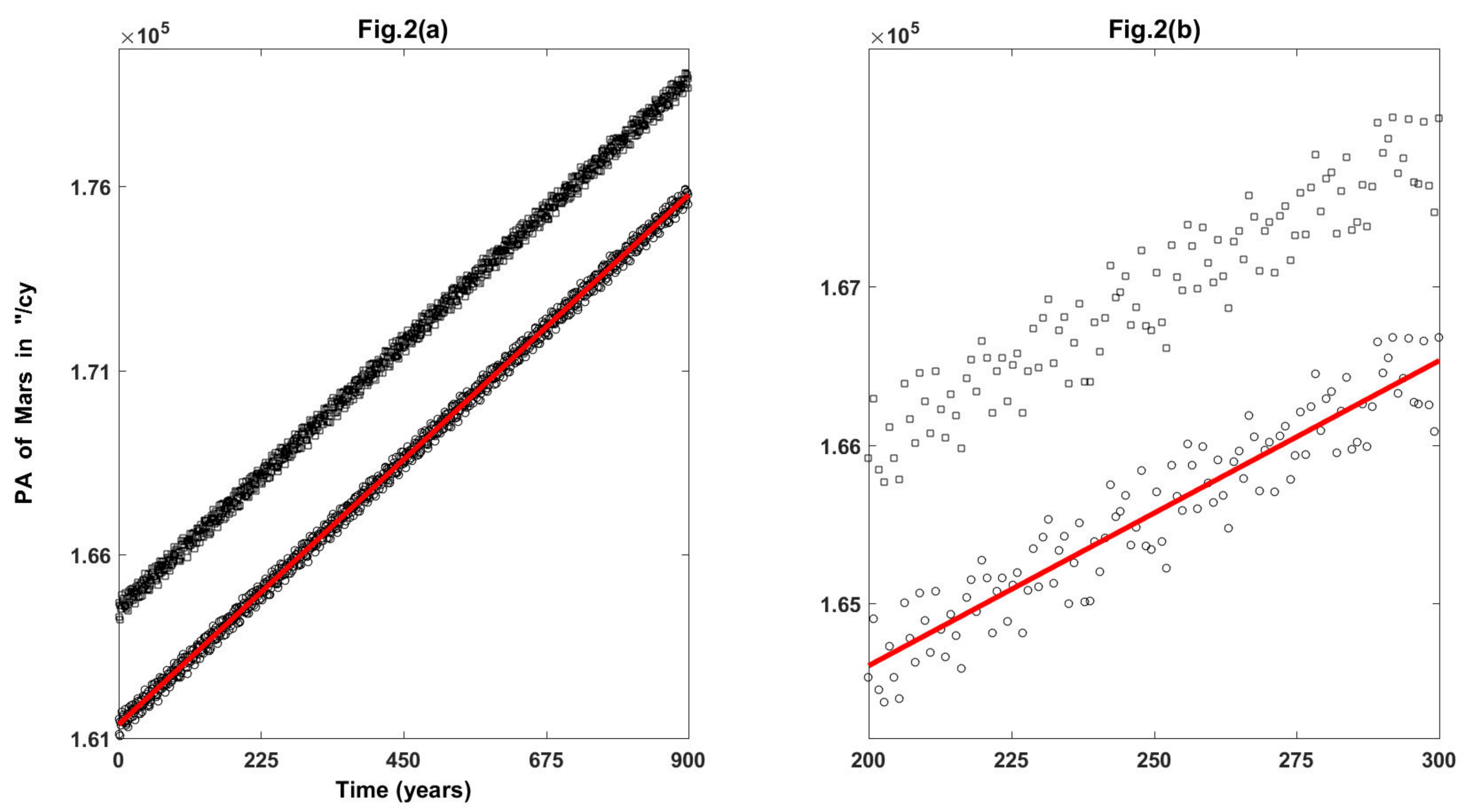}}
	\caption{\label{Fig.2:Timeevolutionmars} Time evolution of Mars’s LRL vector and longitude of perihelion  ($\varpi = \omega + \cos(i) \times \Omega$) in the Newtonian gravitational model, without the hypothetical planet Vulcan. In (a) Rotation of the LRL vector (circles) and longitude of perihelion $\varpi$ over a 900-year interval (squares). In (b) Zoom over a 100-year interval. Solid lines show best-fit linear trends for the LRL vector. For clarity, the origin of the perihelion longitude curve has been shifted upward by an arbitrary amount.The original 8 million-point time series was uniformly subsampled to 1000 points using equidistant intervals, preserving temporal trends while enhancing visual clarity.
	}
\end{figure*}

Fig. 2a illustrates a highly linear evolution of Mars' PA over a 900-year time interval, regardless of the calculation method. The curves exhibit striking similarity and proximity, appearing virtually indistinguishable at the indicated scale. As shown in the zoomed 100-year interval in Fig. 2b, the deviations from the linear trend do not appear to be randomly distributed according to a normal distribution, although such a distribution is assumed in the construction of the 95\% CI. 

Let us now extend the classical Newtonian gravitational framework to incorporate the hypothetical planet Vulcan, defined by a specific set of orbital parameters and mass. Note that the optimization used in the present work is substantially different from that described in my previous works \cite{Pogossian2022},\cite{Pogossian2023}. In previous studies \cite{Pogossian2023},\cite{Pogossian2025}, I optimized the mass and orbital parameters of the hypothetical planet Vulcan, situated within Mercury's orbit, to explain the anomalous Mercury's PA as predicted by GR, all within the framework of the Vulcan hypothesis. Restricting Vulcan's orbit to the intramercurian stability zone prevents the selection of orbital and physical characteristics that would allow for simultaneous corrections to the motions of Mercury, Earth, and Mars.
This study aims to determine Vulcan’s position, mass, and orbital elements by analyzing the gravitational perturbations it would exert on Mercury, Earth, and Mars, in order to identify a set of its orbital parameters that simultaneously reproduces the observed PA of the inner planets, excluding Venus, in a manner consistent with both GR and observational data.

Orbital optimization of the hypothetical planet Vulcan constrains its semi-major axis to \( a \) = 0.545 AU, placing it between Mercury and Venus. This specific configuration is necessary to produce gravitational perturbations on Earth and Mars that are comparable to those predicted by GR. Historically, many astronomers, starting with Le Verrier, hypothesized that the planet Vulcan orbited the Sun between Mercury and the Sun, but despite systematic searches in this region, no confirmed detections have been made\cite{Baum1997}. Very few observations of Vulcan have been carried out in the region between the orbits of Venus and Mercury. Furthermore, at such a distance \( a \)= 0.545 AU, the period of revolution would be 147 days according to Kepler’s third law, which  will significantly affect the expected frequency of transits and expand the search area compared to the traditionally assumed intra-Mercurial orbit.\\ 
Another noteworthy distinction compared to the traditionally hypothesized intra-Mercurial orbit of Vulcan concerns its estimated mass. My optimization results indicate that Vulcan's mass is approximately 0.375 times that of Mercury, which significantly departs from earlier assumptions that attributed to it a mass exceeding that of Mercury. This will certainly make its observation more difficult, and if we accept the possibility that Vulcan is a primordial black hole \cite{Pogossian2025}, its Schwarzchild radius would be only 0.18 mm, making it virtually impossible for direct observation. 
It should be pointed out that as early as 1890, Newcomb proposed the possible existence of a ring of planetoids located just beyond Mercury's orbit and exhibiting low orbital eccentricity to explain the orbital anomalies observed on the inner planets, in the region between Mercury and Venus \cite{Newcomb1895}. According to his estimates, the total mass of such a ring could range from approximately one-third to as little as one-twentieth of Mercury’s mass, depending on its distance from the planet. Our independently derived mass estimate for Vulcan aligns thus quantitatively with the range predicted by Newcomb’s model. According to Newcomb’s estimation, the inclination angle is also nearly equal to that of Mercury’s orbit, which provides additional support for our results.
The PA of Mercury, Earth, and Mars shows significantly reduced sensitivity to the remaining orbital elements of Vulcan. Therefore, the remaining orbital parameters used in this study align with those from the optimized set in my previous research on the Vulcan hypothesis\cite{Pogossian2023}, \cite{Pogossian2025}.  

These orbital elements of Vulcan allow obtaining the LRL vector rotation to be 
$(575.37 \pm 0.00)^{\prime\prime}$/cy for Mercury, $(1160.90 \pm 0.08)^{\prime\prime}$/cy for Earth,  and  $(1592.66 \pm 0.04)^{\prime\prime}$/cy for Mars. In Table-2,  the value of the PA within the framework of classical Newtonian gravitation is determined through two methods: the rotation of the LRL vector, and the time evolution of the perihelion longitude ($\varpi$). Both approaches incorporate the influence of the hypothetical planet Vulcan in the gravitational model. As in Table-1, in a separate column for each method of determining the PA, the coefficient of determination $\mathbf{R^2}$ associated with quality of the linear fit is presented. \\

\begin{table*}[h]
	\centering
	\caption{This table presents the PA determined for each solar system body using two distinct methods: the rotation of the LRL vector and the time evolution of the longitude of perihelion ($\varpi$) in the presence of hypothetical planet Vulcan as in \cite{Pogossian2023},\cite{Pogossian2025}. Each method’s results are shown in separate columns, followed by the corresponding coefficient of determination ($\mathbf{R^2}$) reflecting the quality of the linear fit.}
	\label{tab:Table2}
	\begin{tabular}{|p{2cm}||p{2.5cm}|p{2.5cm}|p{2.5cm}|p{2.5cm}|} 
		\hline
		\multicolumn{5}{|c|}{\textbf{Table 2 :  PA within framework of Vulcan hypothesis}} \\
		\hline
		\textbf{Body} & \textit{PA (LRL)} in $^{\prime\prime}$/cy & $\mathbf{R^2}$ & \textit{PA} ($\varpi$) in $^{\prime\prime}$/cy & $\mathbf{R^2}$ \\
		\hline
		Mercury & $575.37 \pm 0.00$ & 0.99997 & $575.72 \pm 0.00$ & 0.99996 \\ 
		\hline
		Venus & $-7.96 \pm 0.18$ & 0.00091 & $-9.97 \pm 0.18$ & 0.00143 \\
		\hline
		Earth & $1160.90 \pm 0.08$  & 0.99023 & $1158.52 \pm 0.08$ & 0.99019 \\
		\hline
		Mars & $1592.66 \pm 0.04$  & 0.99854 & $1591.98 \pm 0.04$ & 0.99854 \\
		\hline
		Icarus & $339.86 \pm 0.01$  & 0.99715 & $964.12 \pm 0.02$ & 0.99912 \\ 
		\hline
	\end{tabular}
\end{table*}

I observed a notable correlation regarding the influence of Vulcan on the apsidal precession of the inner planets when compared with observational values\cite{Clemence1947}. The PA values for Mercury, Earth, Mars, and Icarus, derived from the slope of the linear fit, exhibit narrow 95\% CI, indicating statistically significant linear trends. The very high coefficient of determination ($\mathbf{R^2}$ > 0.99) confirms the presence of a robust secular linear evolution in the PA of these bodies. \\However, a very important divergence of PAs obtained by two different methods for the asteroid Icarus persists. 
These two methods yield convergent results for PA of  planets but divergent results in the case of the asteroid Icarus. Let now calculate PA of inner planets and Icarus using calculations based on simplified model GR model and described in detail in \cite{Anderson1975}, \cite{Quinn1991},\cite{Benitez2008}.  Total PA can be derived from the equations of motion of a celestial body by inserting relativistic corrections into the Newtonian equations \cite{Anderson1975}, \cite{Quinn1991},\cite{Benitez2008}.  

\begin{table*}[h]
	\centering
	\caption{This table shows the PA determined for each solar system body using two distinct methods: the rotation of the LRL vector and the time evolution of the longitude of perihelion ($\varpi$), calculated using simplified GR as in \cite{Anderson1975}. Each method’s results are shown in separate columns, followed by the corresponding coefficient of determination ($\mathbf{R^2}$) reflecting the quality of the linear fit.}
	\label{tab:Table3}
	\begin{tabular}{|p{2cm}||p{2.5cm}|p{2.5cm}|p{2.5cm}|p{2.5cm}|} 
		\hline
		\multicolumn{5}{|c|}{\textbf{Table 3: PA in the framework of simplified GR model}} \\
		\hline
		\textbf{Body} & \textit{PA (LRL)} in $^{\prime\prime}$/cy & $\mathbf{R^2}$ & \textit{PA} ($\varpi$) in $^{\prime\prime}$/cy & $\mathbf{R^2}$ \\
		\hline
		Mercury & $574.92 \pm 0.00$ & 0.99998 & $577.02 \pm 0.00$ & 0.99998 \\ 
		\hline
		Venus & $-26.81 \pm 0.18$ & 0.01029 & $-26.54 \pm 0.18$ & 0.01008 \\
		\hline
		Earth & $1160.75 \pm 0.08$  & 0.99022 & $1160.71 \pm 0.08$ & 0.99022 \\
		\hline
		Mars & $1593.02 \pm 0.04$  & 0.99854 & $1593.81 \pm 0.04$ & 0.99854 \\
		\hline
		Icarus & $258.41 \pm 0.02$  & 0.99296 & $762.66 \pm 0.02$ & 0.99882 \\ 
		\hline
	\end{tabular}
\end{table*}

In Table 3 values of PA of inner planets and Icarus are given calculated in the frame of simplified GR theory conform to the relativistic contribution calculated in references \cite{Anderson1975}, \cite{Quinn1991},\cite{Benitez2008}. Let us compare the PA of Mercury, Earth, and Mars obtained via the LRL vector rotation method in the classical Newtonian framework (531.97$^{\prime\prime}$/cy, 1156.86$^{\prime\prime}$/cy, and 1591.68$^{\prime\prime}$/cy from Table-1) with the corresponding values computed using the simplified GR model (574.87$^{\prime\prime}$/cy, 1160.75$^{\prime\prime}$/cy,  and 1593.02 $^{\prime\prime}$/cy from Table-3).  Let us also compare the classical Newtonian PA of Mercury, Earth, and Mars respectively derived from time series of perihelion longitude $\varpi$\ (533.38$^{\prime\prime}$/cy, 1155.21$^{\prime\prime}$/cy, and 1591.44 $^{\prime\prime}$/cy from Table-1) gravitation with the corresponding values computed using the simplified GR model (577.02 $^{\prime\prime}$/cy, 1160.71$^{\prime\prime}$/cy, and 1593.81 $^{\prime\prime}$/cy from Table-3). The correspondence of PA of Earth and Mars calculated by both methods are very convincing and exhibit only minor differences. 
However, for Mercury, the discrepancy amounts to 2.1 $^{\prime\prime}$/cy. The LRL value of PA of Mercury is closer to the accepted values of total PA predicted by GR \cite{Will2014_OK}, \cite{Park2017}, \cite{Clemence1947}. Nevertheless, there is only a minor discrepancy when computing the purely relativistic effect for Mercury as the difference between the GR and classical models, we obtain 42.9 $^{\prime\prime}$/cy for the LRL method and 43.64$^{\prime\prime}$/cy for the $\varpi$\ -based approach in close agreement with the theoretical predictions reported by Clemence \cite{Clemence1947} 43.03 $^{\prime\prime}$/cy.\\ 
The relativistic components of the PA for Earth and Mars, calculated using the rotation of the LRL vector, are found to be 3.89 $^{\prime\prime}$/cy and 1.34 $^{\prime\prime}$/cy, respectively. The values obtained for relativistic components of the PA by the method based on the perihelion longitude, yields slightly higher values: 5.50 $^{\prime\prime}$/cy for Earth and 2.37 $^{\prime\prime}$/cy for Mars. The results obtained via the LRL vector method are therefore in closer agreement with the theoretical predictions reported by Clemence \cite{Clemence1947}, who found values of 3.84 $^{\prime\prime}$/cy for Earth and 1.35 $^{\prime\prime}$/cy for Mars. The small discrepancies between methods likely stem from theoretical simplifications in the modeling of relativistic perturbations, methodological differences in computation, and inherent limitations in numerical precision.

Let us now analyze the PA of asteroid Icarus by examining the evolution of the LRL vector and the time series of its perihelion longitude. This analysis is conducted within the framework of three gravitational models, the classical Newtonian gravity model, the GR model, and the Newtonian model with the additional assumption of Vulcan as a perturbing influence.
The GR value of PA based on the perihelion longitude determination from Table-3  is 764.95$^{\prime\prime}$/cy while the Newtonian classical value is only 968.43$^{\prime\prime}$/cy from Table-1.  It means the purely relativistic contribution is -203.48”/c meanwhile the classical Einstein’s formula for purely relativistic  effect is of only about +10 $^{\prime\prime}$/cy \cite{Shapiro1971}. 
This finding contradicts my previously established methodology, which employed two independent methods for evaluating the relativistic contribution to PA. This discrepancy raises concerns regarding the reliability of the $\varpi$\ -based  approach for accurately determining asteroidal PA.\\
Table-3 reveals that the PA value of Icarus calculated by LRL vector rotation within the simplified GR framework is 258.41 $^{\prime\prime}$/cy while the Newtonian classical value is estimated from Table-1 to be 248.48 $^{\prime\prime}$/cy. The difference is about 9.93 $^{\prime\prime}$/cy that is perfectly consistent with the value provided by Einstein’s formula Eq. (1). \\Thus the PA evaluation based on LRL vector rotation seems to be more consistent with Einstein’s formula than the estimation by the help of times series of perihelion longitude. Consequently, I insist on the need for careful interpretation in determining the orbital parameters and long-term evolution of asteroids, as certain inaccuracies in the determination of their orbits can have consequences that go far beyond academic research and directly affect the safety of our planet.\\
Finally let us now compare the GR-predicted value of Icarus’s PA with that obtained under the Vulcan hypothesis, both derived from the determination of LRL vector rotation. As seen above, the LRL vector rotation, calculated either by taking the difference between the GR-corrected and classical Newtonian models or by using Einstein’s formula, amounts to approximately 10$^{\prime\prime}$/cy. 
In the Vulcan hypothesis, Icarus' PA is around  339.87 $^{\prime\prime}$/cy as shown in Table 2, whereas the purely Newtonian PA in Table 1 is only 248.48 $^{\prime\prime}$/cy . Thus, the presence of Vulcan leads to an additional advance of 91.39 $^{\prime\prime}$/cy of Icarus' perihelion, which is nine times greater than that predicted by the relativistic influence of the Sun. It is extremely important to conclude that this difference is measurable by observations of the precession of Icarus' perihelion.\\
In Fig. 3, Icarus' PA is determined within the framework of 3 different theoretical models of gravity. The lowest curve, with the smallest slope, represents the prediction based on classical Newtonian gravity, which, as can be seen from Table 1, is only 248.48 $^{\prime\prime}$/cy. The middle curve shows the observed PA of Icarus, calculated using a simplified GR model\cite{Anderson1975}. It increases more rapidly, indicating an advance of 10 $^{\prime\prime}$/cy faster than predicted by Newtonian theory. The upper curve corresponds to a Newtonian model that includes the additional gravitational attraction of a hypothetical planet Vulcan\cite{Pogossian2023}. This curve has the steepest slope, with a PA of 91.39$^{\prime\prime}$/cy.

\begin{figure*}[!htb]
	\centering
	\resizebox{0.999\textwidth}{!}{\includegraphics[height=3.88 cm, width=7.2 cm,angle=0]{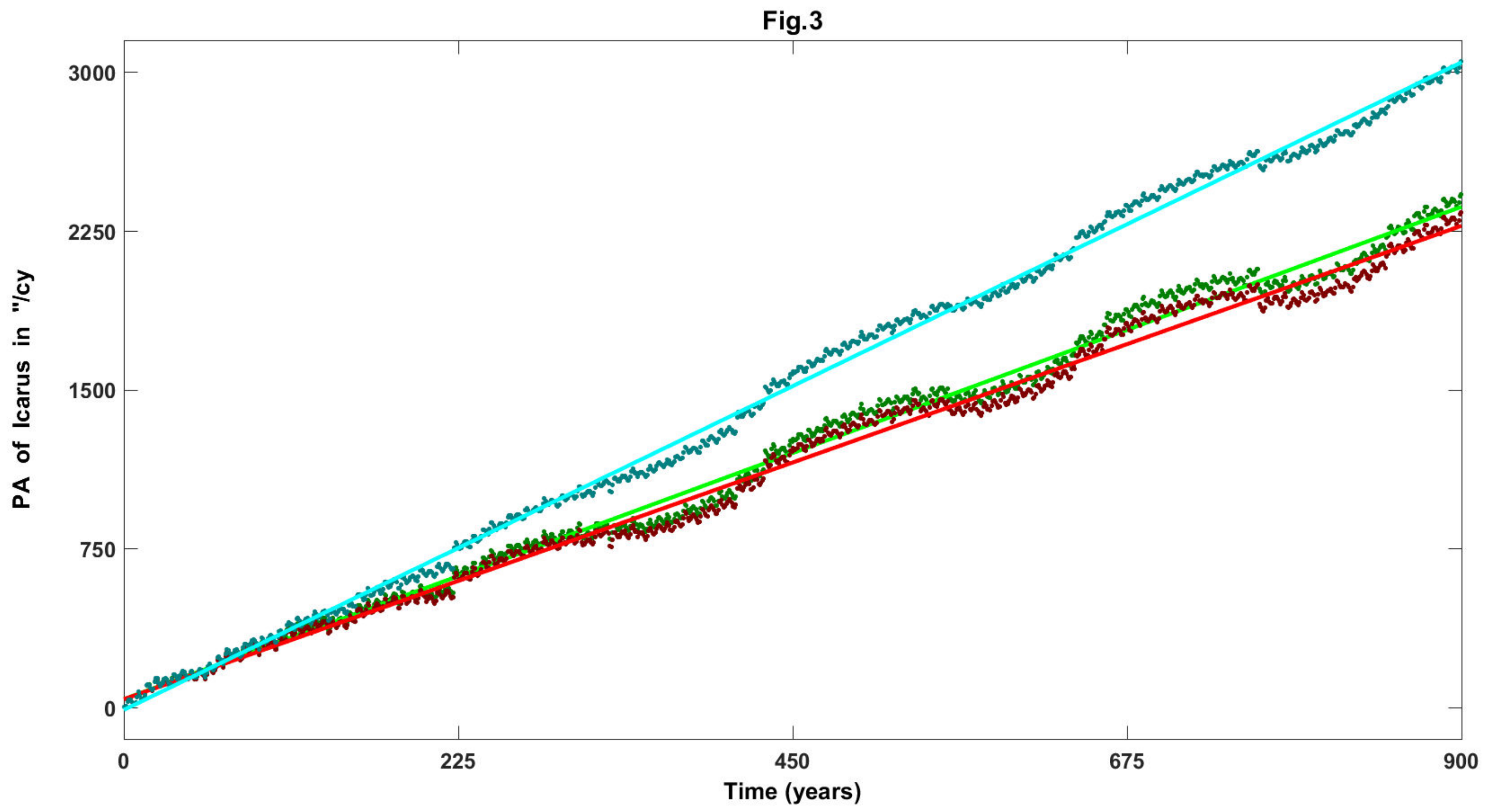}}
	\caption{\label{Fig.3:Time evolutionof PA of Icarus} Time evolution of the PA of Icarus, determined by the rotation of the LRL vector. The lowest curve (in red) represents the Newtonian prediction without the influence of Vulcan. The middle curve (in green) corresponds to a simplified GR model. The uppermost curve (in cyan) shows the Newtonian model including the gravitational pull of the hypothetical planet Vulcan. Solid lines represent linear fits to each respective curve.To improve visual clarity and reduce rendering complexity, the time series data (originally comprising 8 million points) were uniformly downsampled to 1000 points. The subsampling was performed using equally spaced intervals to preserve the overall temporal structure and trends of the signal.
	}
\end{figure*}

Thus, it is important to be able to accurately calculate the revolution of asteroids around the Sun, as their orbits can evolve dynamically and, in the case of near-Earth objects, pose a significant potential threat to our planet. For example, the small deviations of asteroid Apophis from the trajectory predicted by GR underline the importance of high-precision assessments of this high-impact object for planetary defense modeling, particularly in light of its close flyby of Earth in 2029 and its importance for planetary defense strategies.\\ 
From this point of view, the results presented in this work are doubly important. Firstly, it has been demonstrated that the LRL approach gives more consistent results for Icarus PA calculations than that based on times series of perihelion longitude. Secondly, the definitive validation of Anderson's simplified relativity model for the solar system, in comparison with the Vulcan hypothesis, can now be verified by experimental observations.  Therefore, precise observational data of Icarus's orbital dynamics offers a definitive means to either rule out this specific Vulcan hypothesis or, conversely, indicate the necessity of further refinement of GR's application within the Solar System.\\

	\section{Conclusion}
	In this work, I conducted an in-depth comparison of two methods for calculating the PA of inner planets: one based on the rotation of the LRL vector and the other on the time evolution of the perihelion longitude. While both approaches yield nearly identical PA values for the inner planets under classical Newtonian gravity and the GR framework, a significant discrepancy arises in the case of the asteroid Icarus.
	Both methods characterize the PA of inner planets through the slope of a linear trend, which I analyze using linear regression. By treating the deviations from this trend as random, normally distributed error terms, I perform a detailed assessment of the slope's standard deviation and subsequently derive the 95\% CI utilizing the Student’s t-distribution.
	By comparing the PA of Icarus through numerical calculations with the prediction of the same advance by Einstein's formula, I conclude that only the method based on LRL vector rotation provides a coherent explanation for the observed behavior of Icarus' PA. An analysis of the error in the PA evolution slope for Venus allows us to exclude Venus from the list of inner planets considered.
	
	Next, I include a hypothetical planet Vulcan into the Newtonian gravitational model of the solar system and analyze its influence on the PA of the inner planets, using a set of Vulcan parameters obtained through optimization. Optimization constraints enable a finely tuned balance between Vulcan’s mass and semi-major axis, ensuring compatibility with the observed PA of Mercury, Earth, and Mars. For a hypothetical Vulcan with a mass approximately one-third that of Mercury and a semi-major axis of \( a \)=0.545 AU, positioned between Mercury and Venus, would generate gravitational perturbations on Mercury, Earth and Mars consistent with GR predictions. However, unlike a purely relativistic solution, such a Vulcan would exert a far more pronounced effect on Icarus’s PA, enhancing it by roughly a factor of nine compared to GR’s predictions.

	Therefore, accurate observational data of Icarus's orbital dynamics provides a conclusive way to either rule out this particular Vulcan theory or, alternatively, reveal the need for further refinement in the application of General Relativity to the Solar System.


\end{document}